\documentclass[,amsmath,amssymb,floatfix,superscriptaddress
]{revtex4}
\usepackage[utf8]{inputenc}
\usepackage{graphicx}
\usepackage{dcolumn}
\usepackage{bm}
\usepackage{natbib}
\usepackage{epstopdf}
\usepackage{natbib}
\usepackage{amsmath} 
\usepackage[abs]{overpic}
\usepackage[normalem]{ulem}
\usepackage{gensymb}
\usepackage{subfigure}
\usepackage{color} 
\usepackage{tabularx}
\newcommand{\upperRomannumeral}[1]{\uppercase\expandafter{\romannumeral#1}}

\begin{document}
\title{Film coating by directional droplet spreading on fibers}

\author{Tak Shing Chan$^0$}
\affiliation{Department of Mathematics, Mechanics Division, University of Oslo, N-0851 Oslo, Norway.}
\author{Carmen L. Lee$^0$}
\affiliation{Department of Physics and Astronomy, McMaster University, 1280 Main Street West, Hamilton, Ontario, L8S 4M1, Canada.}
\author{Christian Pedersen}
\affiliation{Department of Mathematics, Mechanics Division, University of Oslo, N-0851 Oslo, Norway.}
\author{Kari Dalnoki-Veress}
\affiliation{Department of Physics and Astronomy, McMaster University, 1280 Main Street West, Hamilton, Ontario, L8S 4M1, Canada.}
\affiliation{UMR CNRS Gulliver 7083, ESPCI Paris, PSL Research University, 75005 Paris, France.}
\author{Andreas Carlson}
\email{acarlson@math.uio.no}
\affiliation{Department of Mathematics, Mechanics Division, University of Oslo, N-0851 Oslo, Norway.}

\date{\today}

\begin{abstract}
Plants and insects use slender conical structures to transport and collect small droplets, which are propelled along the conical structures due to capillary action. These droplets can deposit a fluid film during their motion, but despite its importance to many biological systems and industrial applications the properties of the deposited film are unknown. We characterise the film deposition by developing an asymptotic analysis together with experimental measurements and numerical simulations based on the lubrication equation. We show that the deposited film thickness depends significantly on both the fiber radius and the droplet size, highlighting that the coating is affected by finite size effects relevant to film deposition on fibres of any slender geometry. We demonstrate that by changing the droplet size, while the mean fiber radius and the Capillary number are fixed, the thickness of the deposited film can change by an order of magnitude or more. We show that self-propelled droplets have significant potential to create passively coated structures.
\end{abstract}


\maketitle
 \footnotetext[0]{These authors contributed equally.}
\section{Introduction} 
Droplets on slender conical substrates will self-propel due to capillary action~\cite{lorenceau2004,Er2013,chan2020a,chan2020b,Liang2015,McCarthy19,Renvoise2009, Chou2011,jian2007, wu2006} provided the droplets are smaller than the capillary length. This principle is used by insects~\citep{Zheng2010, Parker2001} and plants~\citep{Chen2018,guo2015experimental,Ju2012,Luo2015,Pan2016,Malik2015,shanahan2011,tan2016} for droplet collection. Several studies have focused on mimicking structures found in nature to control liquid movement~\citep{Chen2013,bai2010,cao2014,heng2014,hou2013,ju2013,xu2016}. Recent work~\citep{Chen2018} has shown that the conically shaped {{trichomes}} on the underside of the lid of the {\em{Sarrancenia}}, a pitcher plant, can transport droplets with a velocity several orders of magnitude larger than found in other plants. Enhanced water transport is the result of surface lubrication of the trichome. The first droplet that slowly spreads across the trichome deposits a microscopic liquid film and the following droplets slide along the lubricating film on the pre-wet trichome. From a technological point of view, understanding the principles of film deposition by capillary driven motion of droplets can provide pathways for re-lubrication of slippery liquid infused porous surfaces with conical shapes~\citep{Wong2011,McCarthy19} as well as the development of other multifunctional materials. This lubricating film-coating principle has a fundamental role in biological phenomena and has untapped potential as a droplet-driven coating technique, yet the properties of the liquid film are unknown. We study here how droplets deposit lubricating films as they move along slender structures. 

\begin{figure}
\begin{center}
\includegraphics[width=0.47\textwidth]{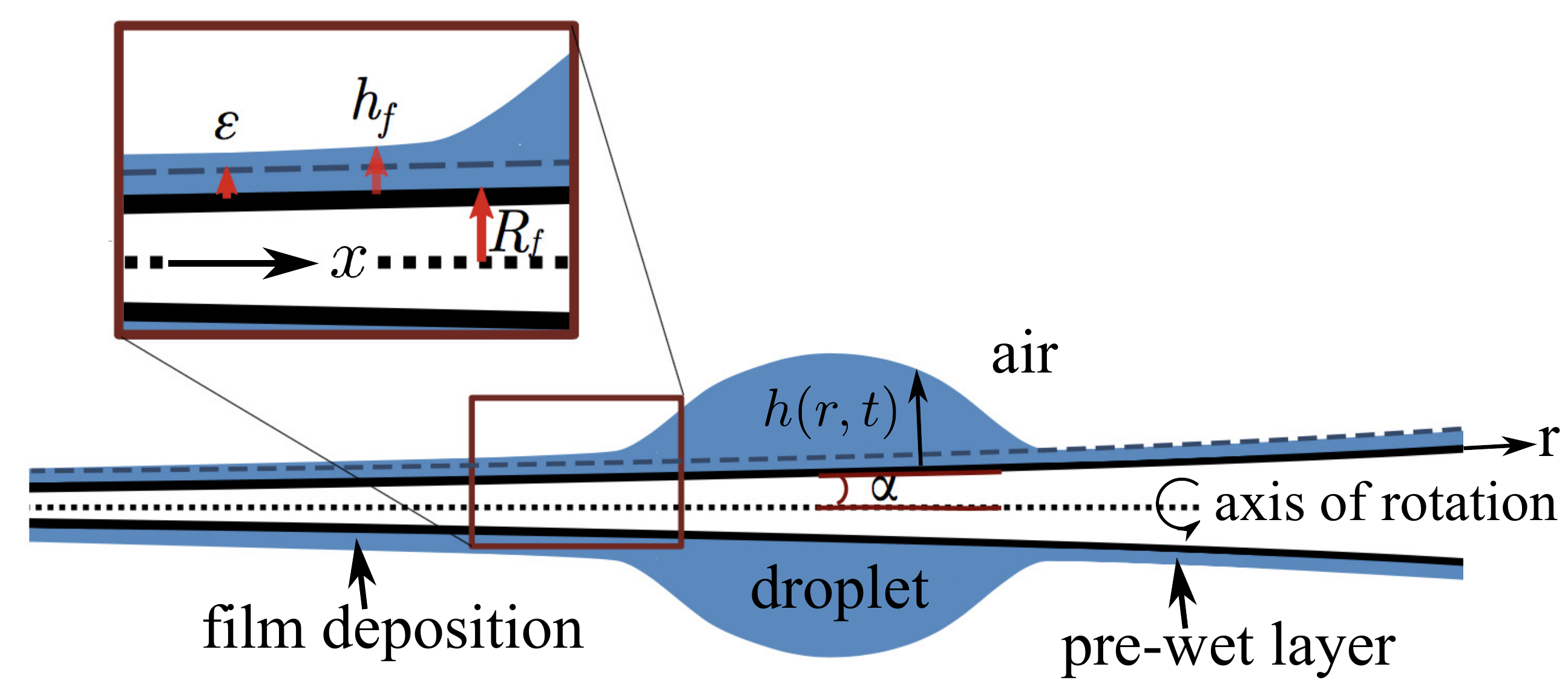}
\caption{(a) A sketch of a droplet on a conical fiber with a local cone angle $\alpha$. Inset: zoom into the region connecting the deposited film of thickness $h_f$ and the receding edge of the droplet at the fiber radius $R_f$. The fiber is pre-wet with a layer of the same fluid of thickness $\epsilon$. }\label{fig1}
\end{center}
\end{figure} 
Coating a solid substrate with a lubricating liquid film as a way to reduce friction between substrates has been known since ancient Egypt~\cite{Dowson98}. The broad relevance of coating processes have made them widely studied  with great advances in understanding their underlying physical principles~\cite{Ruschak1985,Q99,WeRu04,SZAFE08,RIO2017100}. Dip-coating is today one of the most widespread coating techniques~\citep{GBQ03}, where the solid moves with a velocity $U$ relative to the liquid bath. The foundational work of Landau-Levich-Derjaguin (LLD)~\citep{LL42,D43} has paved the way for a fundamental understanding of film deposition on solid substrates during wetting. By considering the viscous capillary flow of a liquid with a viscosity $\mu$ and a surface tension $\gamma$, LLD predicted that the deposited film tickness $h_f$, normalized by the characteristic length of the system $L$, is given by $h_f/L\sim \textrm{Ca}^{2/3}$~\cite{LL42,D43};  where the Capillary number $ \textrm{Ca}\equiv\mu U/\gamma$ is the ratio of the viscous and surface tension forces. The LLD theory was developed for $ \textrm{Ca}\ll1$ and when inertia can be neglected. It is a generic scaling and has proven as an accurate description of a wide range of coating phenomena, i.e. dip coating of plates~\citep{MALEKI2011359}, cylinders~\citep{deryck1996,Q99,Shen2002}, and Bretherthon bubbles~\citep{B61}. However, a droplet depositing a film on a cylinder has a fundamental difference from film deposition from a liquid reservoir; the droplet size introduces another length scale to the system. The fiber geometry and droplet size are tuneable parameters to control the coating process~\cite{lorenceau2004, Er2013}.

\begin{figure}
\begin{center}
\includegraphics[width=0.48\textwidth, angle=0]{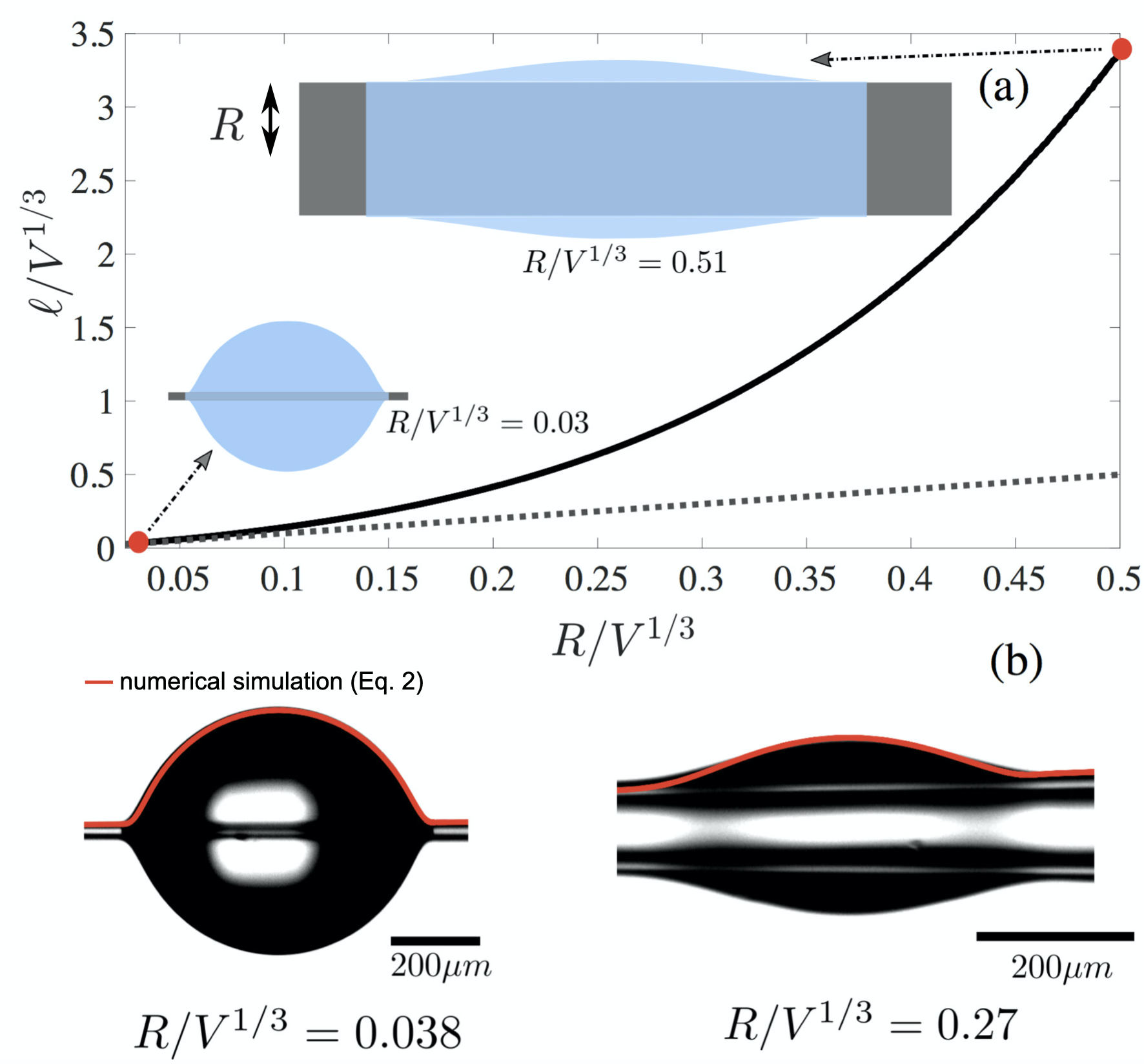}
\caption{(a) The dimensionless characteristic length $\ell/V^{1/3}$ as a function of the rescaled radius $R/V^{1/3}$ of a cylindrial fiber (solid line). The dotted line represents the linear relation, i.e. $\ell=R$. Inset: two static droplets of the same volume in contact with a fiber with $R/V^{1/3}=0.03$ and $R/V^{1/3}=0.51$ (indicated by the two red dots) which demonstrate different droplet shapes when $R$ varies. (b) Sample images of two droplets taken with optical microscopy. Left: $\alpha = 0.03\degree$ and $R/V^{1/3}$ = 0.047 and right:  $\alpha = 2.5\degree$ and $R/V^{1/3}$ = 0.27. The numerically calculated profiles from the lubrication theory on a cone are shown in red for matching $V$, $R$, and $\alpha$.}
\label{fiber_profile}
\end{center}
\end{figure} 

\section{Results}
In the system studied here, a droplet deposits a film as it migrates towards the thicker part of a pre-wet conical fibre, driven by the curvature gradient, as shown schematically in Fig. \ref{fig1}. We investigate the system by combining asymptotic analysis, experiments and numerical simulations. The assumptions made are that there is viscous flow driven  by capillarity ($ \textrm{Ca}\ll 1$). Furthermore, we neglect gravitational effects because the drop size is much smaller than the capillary length, as is clear from the Bond number which  represents the balance between  gravity and surface tension, $ \textrm{Bo}=\Delta \rho gV^{2/3}/\gamma\ll1$; where  $\Delta \rho$ is the density difference between the liquid and surrounding air, $V$ is  the droplet volume,  and $g$ is the gravitational acceleration. As will be seen below, these assumptions are verified by our experiments. 
\subsection{Asymptotic analysis}
We start off by revising the classical LLD theory for the case of a droplet  moving on a cylindrical fibre with radius $R$, by matching asymptotically the quasi-static droplet profile on the fibre $h_s(x)$,  and the self-similar deposited film profile (for details, see \cite{chan2020b}). By matching the profiles, we show that the film thickness $h_f$ scales with $ \textrm{Ca}$ as~\cite{chan2020b}
\begin{equation}\label{hf}
h_f =1.338\ell  \textrm{Ca}^{2/3},
\end{equation}
where $\ell\equiv 1/[\partial^2h_s(x=x_{cl})/\partial x^2]$ is the inverse of the second derivative of the static profile $h_s(x)$ evaluated at the contact line position $x=x_{cl}$, i.e. where the profile $h_s(x)$ meets the solid substrate. A crucial difference to the classical LLD theory is that choosing $\ell=R$ only recovers the correct film thickness in the limit of $R\ll V^{1/3}$. In general, $\ell$ depends on both the droplet volume $V$ and the fiber radius $R$, which indicates a finite size effect.  To illustrate this point, we plot $\ell/V^{1/3}$ as a function of $R/V^{1/3}$ in Fig. \ref{fiber_profile}(a). In the limit where $R\ll V^{1/3}$, $\ell \rightarrow R$ the film thickness $h_f =1.338R \,\textrm{Ca}^{2/3}$ is independent of the droplet volume. However, when $R/V^{1/3}\gtrsim 0.15$ the droplet size starts to play a significant role in predicting the deposited film thickness that is much larger than if we would na\"ively assume $\ell=R$. Since $\ell/V^{1/3}$ increases with $R/V^{1/3}$ faster than a linear relation, Eq. (\ref{hf}) also implies that a smaller droplet deposits a thicker film for fixed $R$ and $ \textrm{Ca}$. For the directional spreading of droplets on a conical fibre with a small cone angle $\alpha$, the influence of $\alpha$ on $\ell$ only appears as high order corrections which are neglected here (see \cite{chan2020b} for details). The conical geometry acts as a factor that generates the spontaneous motion of the droplet and plays a role in determining the magnitude of $ \textrm{Ca}$. The theoretical prediction for the film thickness dependence on droplet size [see Eq. (\ref{hf})] can now be compared to experiments and numerical simulations based on the lubrication theory. 

\subsection{Experiments}
The conical substrates used in the experiments are prepared by pulling standard borosilicate glass capillary tubes in a magnetic micropipette puller (Narishige PN-30). The resulting shape of the capillary tube is a nearly conical fiber with a smoothly varying diameter and gradient, with a smaller cone angle nearing the tip of the fiber. The gradient in the cone angle varies slowly along the fiber, thus on the length scale of the droplet the fibers can be treated as ideal. Droplets of silicone oil with viscosity of $\mu\approx 4.9$~Pa$\cdot$s, and with air-liquid surface tension $\gamma$ = 22 mN/m, were deposited at the fiber tip. Silicone oil is ideal because it is totally wetting, chemically stable, non-volatile, and non-hygroscopic. The fiber is pre-wet by placing a droplet on the tip of the fiber and allowing it to migrate from one end of the fiber to the other, thereby depositing a film. Pre-wet film thicknesses were found to range from 0.27 - 13.87 $ \mu$m, as determined by optical microscopy (OM). OM images of the fiber before and after coating are taken and used to obtain the film thickness. Droplets of volumes $V$ in the range of 0.009 - 1.99 mm$^3$ i.e. Bo $\in[0.02 - 0.7]$, were deposited onto the fiber. Images of the droplet are taken as it migrates along the fiber at a given radius $R$, and the deposited film is observed as the droplet passes a given location. Deposited film thicknesses were measured in the range of 0.17 - 19.75 $\mu$m. 

\begin{figure}
\begin{center}
\includegraphics[width=0.5\textwidth]{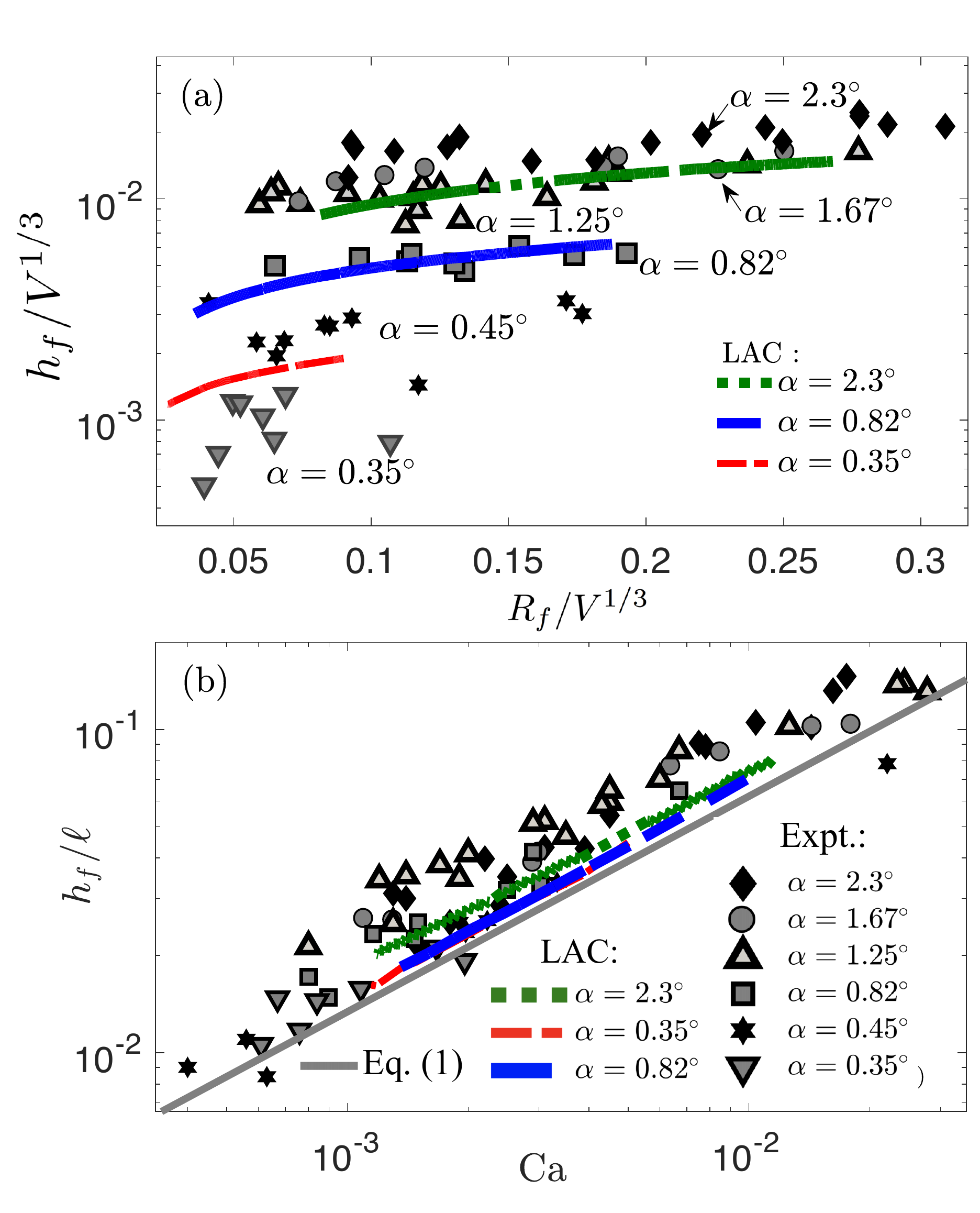}
\caption{(a) The deposited film thickness $h_f$ as a function of the cone radius $R_f$ and the cone angle $\alpha$. Both axis are normalized by $V^{1/3}$. Symbols are experimental data and curves are numerical results from the lubrication theory on a cone (LAC). The pre-wet  layer thickness $\epsilon$ in both the experiment and the theory is controlled within a range of $10^{-2}$-$10^{1}$ $\mu$m. (b) The film thickness $h_f$ rescaled by $\ell$ as a function of the capillary number $ \textrm{Ca}$. The solid line is the result of asymptotic matching given by Eq. (\ref{hf}). \label{Fig_all}}
\end{center}
\end{figure}

\subsection{Numerical simulations}
To give a mathematical description of the droplet flow on the pre-wet  fibre, we turn to the lubrication approximation for the viscous incompressible flow, when the cone angle $\alpha\ll1$. The thin film equation is obtained by reducing the Navier Stokes equations for flow in films with large lateral dimensions in relation to the thickness~\citep{Batch67,Oron97}, in combination with mass conservation. A detailed derivation of the lubrication approximation on a conical geometry for $\alpha\ll1$ is found in~\cite{chan2020a}. Note that we impose a no-slip condition at the solid substrate and no-shear stress at the free surface. The axisymmetric liquid-air interface profile is given by $h=h(r,t)$, defined as the distance between the interface and the substrate, as a function of the radial distance from the vertex of the cone $r$ and time $t$. The evolution of the free surface is described by~\cite{chan2020a,chan2020b},
\begin{eqnarray}\label{lac1}
\frac{\partial h}{\partial t}+\frac{1}{r\alpha+ h}\frac{\partial}{\partial r} \left(M \frac{\partial p}{\partial r}\right)=0,
\end{eqnarray}
where  the mobility $M=M(h,r,\alpha)$ reads
\begin{eqnarray}
M(h,r,\alpha)=&\frac{r^4\alpha^4}{2\mu}\Bigg\{\frac{1}{8}\left[3\left(1+\frac{h}{r\alpha}\right)^4-4\left(1+\frac{h}{r\alpha}\right)^2+1\right] 
-\frac{1}{2}\left(1+\frac{h}{r\alpha}\right)^4\ln(1+\frac{h}{r\alpha})\Bigg\}. 
\label{G}
\end{eqnarray}
The capillary pressure gradient in the liquid generates the flow and the pressure $p=p(r,t)$ reads

\begin{equation}\label{cur}
p=-\gamma\Bigg\{\frac{\frac{\partial^{2} h}{\partial r^{2}}}{\left[1+\left(\frac{\partial h}{\partial r}\right)^{2}\right]^{3/2}}-\frac{1-\alpha \frac{\partial h}{\partial r}}{(r\alpha 
	+h )\left[1+\left(\frac{\partial h}{\partial r}\right)^{2}\right]^{1/2}}\Bigg\},
\end{equation}
where the expression is simplified for $\alpha\ll 1$ \cite{chan2020a,chan2020b}. Eq. (\ref{lac1}) and (\ref{cur}) are discretized by linear elements and numerically solved with a Newton solver by using the open source finite element code FEniCS\citep{logg2012automated}, additional details about the numerical approach are found in~\cite{chan2020b}. The initial condition is a droplet smoothly connected to a pre-wet  film of thickness $\epsilon$.  At the two boundaries ($\delta\Omega$) of the numerical domain we impose $h(\delta\Omega,t)=\epsilon$ and $p(\delta\Omega,t)=\gamma/R(\delta\Omega)$, where $R(\delta\Omega)$ is the radius of the cone at the boundaries. We note that only the droplet volume $V$ is important and the initial droplet shape does not affect the results.

\section{Discussion and Conclusion}
We start by comparing the droplet spreading dynamics on two cones with $\alpha = 0.03\degree$ and $\alpha = 2.5\degree$, where the droplet quickly relaxes from its initial condition to its quasi-static shape and then starts to translate to the thicker part of the fiber. When we overlay the experimental measurement with the numerical simulations, as shown in Fig. \ref{fiber_profile}(b), we see that the two results are in  close agreement. By zooming into the trailing edge of the droplet, both the experiment and the numerical simulation show the deposition of a film of different thickness from that of the pre-wet film $\epsilon$.

Next we turn to characterize the thickness of the film during the droplet spreading dynamics on the fibre. To determine the $ \textrm{Ca}$ number, we extract the droplet velocity $U$ measured at its center of mass. The film is measured on the cone after the droplet has deposited the film, which is stable throughout the observation time in the experiments and the numerical simulations. Since there is a slight gradient in the cone angle along $r$ in the fibre used in the experiments, we extract the cone angle locally at a given position on the cone with radius $R=R_f$, here $R_f$ is the cone radius in the receding region, defined based on the droplet profile see \cite{chan2020b}. The deposited film thickness $h_f$ is then a function of $\alpha$, $R_f$ and $\epsilon$. We combine all the experimental measurements and the numerical predictions of $h_f\in [0.17 - 19.75]$ $\mu$m, i.e. $\alpha\in [0.35-2.3]^{\circ}$, $\epsilon$  in Fig. \ref{Fig_all}a, which are in good agreement. The film thickness is not uniform along the fiber for a fixed cone angle, but increases with the cone radius $R_f$. The film thickness $h_f$ increases by roughly one order of magnitude when the cone angle $\alpha$ is varied from $0.35^{\circ}$ to $2.3^{\circ}$. 

To further compare the theory to the experiments and numerical simulations, we rescale our measurements according to Eq. (\ref{hf}) and also plot the analytical prediction, see Fig. \ref{Fig_all}b. Since the motion of the droplet is driven by capillarity, i.e. it is self-propelled, the droplet velocity is a function of the position on the cone. The deposited film thickness $h_f$ rescaled by $\ell$ obtained from the experiments and the lubrication theory on a cone is shown as a function of $ \textrm{Ca}$ in Fig. \ref{Fig_all}b. When comparing the results (\ref{hf}) predicted by the asymptotic matching, the experiments and the numerical simulations we observe that they are in close agreement, especially on the smallest cone angles. When $\alpha$ increases, there is a slight deviation from $2/3$ scaling observed in the numerical simulations with a slightly larger film thickness than predicted from Eq. 1, likely a consequence of the reduced separation of length scales between the film thickness $h_f$ and the droplet size $V^{1/3}$.

We show that self-propelled droplets have significant potential to create passively coated structures. By combining an asymptotic analysis, experiments and numerical simulations of the lubrication equation, we have demonstrated that a droplet that moves on a fibre can deposit a film with a thickness $h_f$, controlled by the droplet's capillary number and the characteristic length $\ell$. The quantity $\ell$ is a geometric factor which is linear with respect to the fiber radius $R$ when $R/V^{1/3}\ll1$, i.e. the droplet is much greater in size than the fibre radius. Otherwise, $\ell/V^{1/3}$ increases significantly with $R/V^{1/3}$ when $R/V^{1/3}\gtrsim 0.15$.  Our finding has direct implications for control of film deposition during spreading, e.g. if we fix the fiber radius, decreasing the droplet size can increase the thickness of the deposited film by an order of magnitude or more at the same $ \textrm{Ca}$. Coating by droplets introduces novel design features that does not exist in classical coating techniques where the substrate is connected to a liquid reservoir. For a droplet moving on a cylindrical fiber driven by external forces, e.g. electric, magnetic, gravitational, the deposited film thickness follows Eq. (\ref{hf}), whereas $\textrm{Ca}$ depends on the magnitude of the driving force. Our findings are expected to be  relevant for any droplet coating application involving a slender geometry and may help shed light onto why slender conical structures have evolved in a diverse set of biological systems to facilitate efficient droplet transport.

\subsection*{Acknowledgements}
T.C. and A.C. gratefully acknowledge financial support from the Research Council of Norway (project number 301138) and the UiO:LifeScience initiative at the University of Oslo. CLL and KDV acknowledge financial support from the Natural Science and Engineering Research Council of Canada. \\

Declaration of Interests. The authors report no conflict of interest.

\bibliographystyle{apsrev4-1}
\bibliography{new_all_ref-2}
\end{document}